\documentclass[12pt]{article}
\title{\vspace{-2 cm} Magnetic square lattice with vertex coupling of a preferred orientation}

\usepackage{array}
\usepackage{appendix}
\usepackage{graphicx}
\usepackage{amssymb}
\usepackage[T1]{fontenc}
\usepackage[utf8]{inputenc}
\usepackage[raggedright]{titlesec}
\usepackage{blindtext}
\usepackage{commath}
\usepackage{cite}
\usepackage{caption}
\usepackage[textwidth = 7in]{geometry}
\usepackage{mathtools}
\usepackage[dvipsnames]{xcolor}
\usepackage{enumitem}
\usepackage{amsmath}
\usepackage{amsthm}
\usepackage{thmtools}
\usepackage{hyperref, cleveref}
\usepackage{setspace}
\titleformat{\paragraph}[hang]{\normalfont\normalsize\bfseries}{\theparagraph}{1em}{}
\titlespacing*{\paragraph}{0pt}{3.25ex plus 1ex minus .2ex}{0.5em}

\makeatletter
\DeclareRobustCommand{\change}{%
	\@bsphack
	\leavevmode
	\color{magenta}%
	\@esphack
}
\DeclareRobustCommand{\stopchange}{%
	\@bsphack
	\normalcolor
	\@esphack
}
\makeatother

\newcommand{\e}{\mathrm{e}}
\newcommand{\D}{\mathrm{d}}

\stepcounter{secnumdepth}
\stepcounter{tocdepth}

\pagebreak

\pagebreak

\author{Marzieh Baradaran$^1$, Pavel Exner$^{2,3}$, and Ji\v{r}\'{\i} Lipovsk\'{y}$^1$}

\date{\small 1) Department of Physics, Faculty of Science, University of Hradec Kr\'alov\'{e}, Rokitansk\'eho 62, 500 03 Hradec Kr\'alov\'{e}, Czechia \\
2) Doppler Institute for Mathematical Physics and Applied Mathematics, Czech Technical University, B\v rehov\'a 7, 11519 Prague, Czechia \\
3) Department of Theoretical Physics, Nuclear Physics Institute, Czech Academy of Sciences, 25068 \v{R}e\v{z} near Prague, Czechia \\
\emph{marzie.baradaran@yahoo.com, exner@ujf.cas.cz, jiri.lipovsky@uhk.cz}}

\begin{document}

	\captionsetup[figure]{labelfont={bf},labelformat={default},labelsep=period,name={Fig.},font={normalsize}}
	\captionsetup[table]{labelfont={bf},labelformat={default},labelsep=period,name={Table}}
	\maketitle

\begin{abstract}
We analyze a square lattice graph in a magnetic field assuming that the vertex coupling is of a particular type violating the time reversal invariance. Calculating the spectrum numerically for rational values of the flux per plaquette we show how the two effects compete; at the high energies it is the magnetic field which dominates restoring asymptotically the familiar Hofstadter's butterfly pattern.
\end{abstract}

\section{Introduction}
\label{sect:Intro}

Lattice quantum graphs exposed to a homogeneous magnetic field are among systems exhibiting interesting and highly nontrivial spectral properties the origin of which lies in the (in)commensurability of the two inherent lengths, the lattice spacing and the cyclotronic radius, as first noted by Azbel \cite{Az64} and made widely popular by Hofstadter \cite{Ho76}. The setting of the problem may differ: quantum (metric) graphs are through the duality \cite{vB85, Ca97, Ex97, Pa12} related to discrete lattices \cite{Sh94}, and those in turn can be translated into the Harper (or critical almost Mathieu) equation. In particular, the Cantor nature of the spectrum for irrational flux values, the so-called Ten Martini Problem, was one of the big mathematical challenges for two decades. It was finally demonstrated by Avila and Jitomirskaya \cite{AJ09} and more subtle properties of the spectrum were subsequently revealed, see, e.g., \cite{LS16, HLQZ19}. Similar effect was observed in one-dimensional arrays with a magnetic field changing linearly along them with an irrational slope \cite{EV17}.

Quantum graphs \cite{BK13} are most often investigated under the assumption that the wave functions are continuous at the graph vertices, in particular, having the coupling usually called Kirchhoff. This is, however, by far not the only possibility \cite{KS99}, and neither the only interesting one. Following the attempt to model the anomalous Hall effect using lattice quantum graphs \cite{SK15, SV22} it was noted, and illustrated on a simple example, that vertex coupling itself may be a source of a time-reversal violation \cite{ET18}. The vertex coupling in question was found to have interesting properties, among them the fact that the transport properties of such a vertex at high energies depend crucially on its parity which was shown to lead to various consequences \cite{EL19, BET20, BET22, BE22}. It also exhibited a nontrivial $\mathcal{PT}$-symmetry although the corresponding Hamiltonians were self-adjoint \cite{ET21}.

Since the magnetic field is a prime source of time-reversal invariance violation, in graphs with the indicated coupling we have two such effects that may either enhance mutually or act against each other. The aim of this paper is to analyze a model of a magnetic square lattice graph to see how the preferred-orientation vertex coupling can influence its spectral properties. We will compute the spectrum numerically for various rational values of the magnetic flux; the result will show that at high energies the dominating behavior comes from the field alone. The vertex coupling effects are suppressed asymptotically, however, they never completely disappear. 

The paper is structured as follows. First, we recall how the Hamiltonians of magnetic quantum graphs look like in general (Section~\ref{sect:mggraph}). Then, in Section~\ref{sect:Model}, we investigate in detail the cases when the `unit-flux cell' contains two and three vertices (Subsections~\ref{sect:fluxpi} and~\ref{sect:fluxq3}) and give the properties of the general model (Subsection~\ref{sect:fluxpq12}).

\section{Magnetic graphs with preferred-orientation coupling}
\label{sect:mggraph}

In the usual setting \cite{BK13} we associate with a metric graph a state Hilbert space consisting of (classes of equivalence of) $L^2$ functions on the edges of the graphs. In presence of a magnetic field, the particle Hamiltonian acts as the magnetic Laplacian on it, namely as $\left(-i\nabla-\mathbf{A}\right)^2$ on each graph edge. Here we naturally employ the rational system of units $\hbar=2m=1=e=c$; should a purist object that the fine structure constant does not equal one, we include the corresponding multiplicative factor into the magnetic intensity units. In other words, since the motion on the graph edges is one-dimensional, the operator action on the $j$th edge is $\psi_{j}\mapsto -\mathcal{D}^{2}\psi_{j}$, where $\mathcal{D}$ is the quasi-derivative operator
\begin{equation}\label{quasi-derivative}
	\mathcal{D}:=\frac{\D}{\D x}- i\, A_{j}
\end{equation}
with $A_{j}$ being the tangential component of a vector potential referring to the given field on that edge; as usual we are free to choose the gauge which suits our purposes.

To make such a magnetic Laplacian a self-adjoint operator, the functions at each vertex $v$, connecting $n$ edges, have to be matched through the coupling conditions \cite{KS03}
\begin{equation}\label{eq:coupU}
	(U_v-I_v)\Psi_v + i \ell(U_v+I_v)(\mathcal{D}\Psi)_v = 0,
\end{equation}
where $\ell\in \mathbb{R}_+$ is a parameter fixing the length scale; we set $\ell=1$ here for the sake of simplicity. Furthermore, $U_v$ is an $n\times n$ unitary matrix, $\Psi_v$ and  $(\mathcal{D}\Psi)_v$ are, respectively, vectors of boundary values of the functions $\psi_j(x)$ and their quasi-derivatives, the latter being conventionally all taken in the outward direction. This is a large family; its elements can be obviously characterized by $n^2$ real parameters. The couplings with the wave function continuity, the so-called $\delta$-couplings, form a one-parameter subfamily in it.

Here we are going to consider a particular, very different case of matching condition introduced in \cite{ET18}, which corresponds to a matrix $U_v$ of the circulant type, namely
\begin{equation}
U_v = \begin{pmatrix}
	0 & 1 & 0 & \dots & 0 & 0\\
	0 & 0 & 1 & \dots & 0 & 0\\
	\vdots & \vdots &\vdots & \ddots & \vdots & \vdots\\
	0 & 0 & 0 & \dots & 0 & 1\\
	1 & 0 & 0 & \dots & 0 & 0\\
\end{pmatrix}\,.\label{eq:uv}
\end{equation}
Writing the condition \eqref{eq:coupU} with this matrix $U_v$ in components, we get
\begin{equation}\label{coupB}
	(\psi_{j+1}-\psi_{j})+i \left(\mathcal{D}\psi_{j+1}+\mathcal{D}\psi_{j} \right)=0,\quad\; j=1,\dots,n,
\end{equation}
where $\psi_j$ is the boundary value of the function $\psi_j$ at the vertex, and similarly for $\mathcal{D}\psi_{j}$. The index $j$ labels the edges meeting at the vertex and the numeration is cyclic; we identify $\psi_{n+k}$ with $\psi_{k}$ for $k\in\mathbb{Z}$. This coupling violates the time-reversal invariance exhibiting a preferred orientation, most pronounced at the momentum $k=\ell^{-1}$; to see it, it is enough to recall that $U_v$ is nothing but the on-shell scattering matrix $S(\ell^{-1})$ \cite{BK13}. The violation is also related to the fact that the above matrix $U_v$ is manifestly non-invariant with respect to transposition \cite{ET21}.

\section{The model}
\label{sect:Model}

After this preliminary, we can describe the system of our interest in more details. We consider a square lattice, which is placed into a homogeneous magnetic field $\mathbf{B} = (0, 0, B)$ perpendicular to the lattice plane; without loss of generality; we again opt for simplicity and assume that the edges of the lattice cells are of unit length.

We focus on situations which can be treated by methods devised for periodic systems, thus we suppose that the magnetic flux $\Phi$ per plaquette is a rational multiple of the flux quantum which in the chosen units is $\Phi_0=2\pi$, or more specifically, that the dimensionless flux ratio $\frac{\Phi}{\Phi_0} $ is equal to a rational number $ 0<\frac pq<1$ with coprime positive integers $q\geq2$ and $p=1,2,...,q-1$. Parts of the lattice corresponding to the unit flux consist thus of $q$ plaquettes.

\begin{figure}[h]
	\centering
	\includegraphics[scale=0.85]{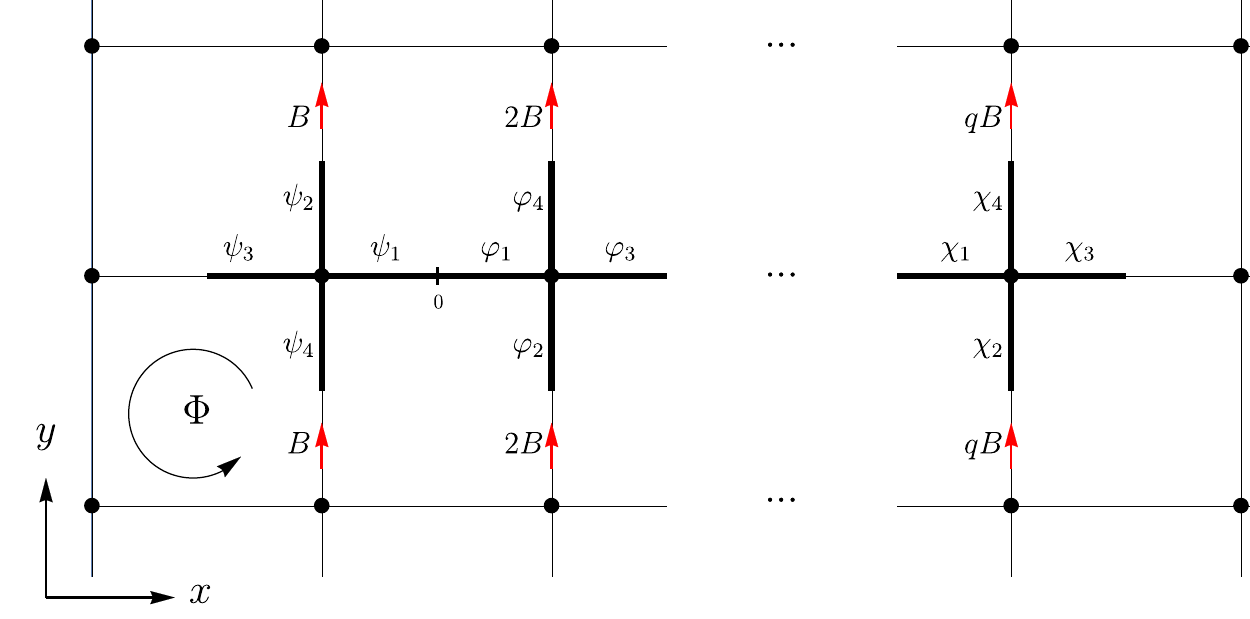}
	\caption{ An elementary cell of the square lattice in a homogeneous magnetic field with the flux value $\Phi=2\pi\frac pq$ per plaquette, consisting of the neighborhood of $q$ vertices of degree four indicated by thick black lines; the arrows on the vertical edges represent the vector potential.}
	\label{unitcaellmag}
\end{figure}

To find the spectrum using the Floquet-Bloch decomposition theorem \cite[Chap.~4]{BK13} we have to choose therefore a plane tiling by domains the areas of which are $q$. Naturally, this can be done in different ways; we choose the simplest one in which the `tiles' are arrays of $q$ elementary cells. In that case, it is suitable to adopt the Landau gauge $\mathbf{A} = B(0, x, 0)\,$ so that $A_x=0$ holds on the horizontal edges, while $A_y$ on the vertical edges is linear with the slope being an integer multiple of $B$. Another ambiguity concerns the choice of the elementary cell; since the main object of our interest is the lattice, we select for it the symmetric cross-shaped neighborhood of a vertex. Consequently, the `unit-flux cell' will contain $q$ vertices of degree four as indicated in Fig.~\ref{unitcaellmag}.

As shown in the figure, the coordinates are supposed to increase `from left to right' and `from bottom to top', in which case the constant components of $A_j$ in \eqref{quasi-derivative} have positive signs on the vertical edges and the magnetic Laplacian operator at such an edge of the magnetic unit cell acts as $-\mathcal{D}_v^{2}:=-\big( \frac{d}{dy}- i v B \big)^{2}$, where $v=1,2,...,q$ is the vertex index. At the horizontal edges, on the other hand, the constant components of $A_j$ in \eqref{quasi-derivative} are absent and the operator acts as the usual Laplacian i.e. $-\frac{d^2}{dx^2}$.

The fiber operators in the Floquet-Bloch decomposition have a purely discrete spectrum; each component of the eigenfunctions with energy $E=k^2>0$ is a linear combination of the functions $\e^{\pm ikx}$ on the horizontal edges, and $\e^{i v B y}e^{\pm iky}$ on the vertical edges with the vertex index $v$. As for the  negative spectrum, the solutions are combinations of real exponentials; one can simply replace $k$ by $i\kappa$ with $\kappa>0$ as we will do in Secs.~\ref{sect:Negatq2} and \ref{sect:Negatq3} below. The fiber operators are labeled by the quasimomentum components $\theta_1,\theta_2\in[-\pi,\pi)$ which indicate how is the phase of the wavefunctions related at the opposite ends of the unit-flux cell; it varies by $\exp (i \theta_1)$ in the $x$-direction, over $q$ elementary cells, and by $\exp(i \theta_2)$ in the $y$-direction, or a single cell size.

To find the spectra of the fiber operators for a given flux value of $\Phi=2\pi\frac pq$ per plaquette as functions of the quasimomentum, and through them the spectral bands of the system, is a straightforward but tedious procedure which becomes increasingly time-consuming as $q$ becomes a larger number. This follows from the fact that the vertices are of degree four, and consequently, to derive the spectral condition, one has to deal for a given $q\geq2$ with a system of $4q$ linear equations, requiring the corresponding $4q \times 4q$ determinant of such a system to vanish.

In this paper, we investigate and report the results concerning all the coprime ratios $\frac{\Phi}{\Phi_0}=\frac pq$ with $q\in\{2,...,12\}$ and $p=1,...,q-1$, having used Wolfram Mathematica 12 for all the calculations. In the next two sections, we describe in detail the derivation of spectral properties of the first two cases, $q=\{2,3\}$, corresponding to the flux values of $\Phi=\pi$ and $\Phi= 2\pi\tfrac{p}{3}$ per plaquette, where $p=\{1,2\}$. For higher values of $q$ the determinants are obtained in a similar way; they are explicit but increasingly complicated functions and we report here just the resulting spectral bands rather than the huge expressions that determine them.

\subsection{The case $\Phi=\pi$}
\label{sect:fluxpi}

We begin with the simplest nontrivial case, $q=2$, and specify the Ans\"{a}tze for the wavefunction components indicated in Fig.~\ref{unitcaellmag} as follows:
\begin{align}\label{ansatzB}
	&\psi_{j}(x)=a_{j}^{+}\e^{ikx}+a_{j}^{-}\e^{-ikx},      & x&\in[-\tfrac{1}{2} ,0]  , \nonumber \\
	&\psi_{2}(y)=\big(a_{2}^{+}\e^{iky}+a_{2}^{-}\e^{-iky}\big)\e^{iBy},        & y&\in[0,\tfrac{1}{2} ]  , \nonumber \\
	&\psi_{4}(y)=\big(a_{4}^{+}\e^{iky}+a_{4}^{-}\e^{-iky}\big)\e^{iBy},      & y&\in[-\tfrac{1}{2} ,0]  ,\nonumber \\[7pt]
	&\varphi_{j}(x)=b_{j}^{+}\e^{ikx}+b_{j}^{-}\e^{-ikx},                       & x&\in[0,\tfrac{1}{2}] , \\
	&\varphi_{2}(y)=\big(b_{2}^{+}\e^{iky}+b_{2}^{-}\e^{-iky}\big)\e^{2iBy},  & y&\in[-\tfrac{1}{2} ,0] , \nonumber \\
	&\varphi_{4}(y)=\big(b_{4}^{+}\e^{iky}+b_{4}^{-}\e^{-iky}\big)\e^{2iBy},   & y&\in[0,\tfrac{1}{2}] , \nonumber
\end{align}
where $j=1,3$; while the $y$-coordinate is the same in both the elementary cells, from computational reasons we consider the range of the $x$-coordinate for each edge segment separately. The functions $\psi_1$ and $\varphi_1$ have to be matched smoothly at the midpoint of the edge; together with the conditions imposed by the Floquet-Bloch decomposition at the endpoints of the unit-flux cell, we have
\begin{align}\label{FloqB}
\psi _1(0)&=\varphi _1(0),       &  \psi _1'(0)&=\varphi _1'(0),  \nonumber\\
	\varphi_3\big(\tfrac{1}{2} \big)&=\e^{i \theta_1 }\, \psi _3\big(\!-\!\tfrac{1}{2} \big),
	& \varphi_3'\big(\tfrac{1}{2} \big)&=\e^{i \theta_1 } \,\psi _3'\big(\!-\!\tfrac{1}{2} \big),\nonumber \\
	\psi _2\big(\tfrac{1}{2} \big)&=\e^{i \theta_2 }\, \psi _4\big(\!-\!\tfrac{1}{2} \big),
	& \mathcal{D}_{1}\psi _2\big(\tfrac{1}{2} \big)&=\e^{i \theta_2 } \,\mathcal{D}_{1}\psi _4\big(\!-\!\tfrac{1}{2} \big),\\
	\varphi _4\big(\tfrac{1}{2} \big)&=\e^{i \theta_2 } \,\varphi _2\big(\!-\!\tfrac{1}{2} \big),
	& \mathcal{D}_{2}\varphi _4\big(\tfrac{1}{2} \big)&=\e^{i \theta_2 }\, \mathcal{D}_{2}\varphi _2\big(\!-\!\tfrac{1}{2} \big),\nonumber
\end{align}
where the operators $\mathcal{D}_{1}$ and $\mathcal{D}_{2}$, as already introduced above, are $\mathcal{D}_v:= \frac{d}{dy}- i v B $ with $v\in\{1,2\}$. Next, imposing the matching conditions \eqref{coupB} at the two vertices of the `magnetic unit cell', and taking into account that the derivatives have to be taken in the outward direction, we arrive at the following set of equations
\begin{align}\label{conditionB}
	& \psi _2(0)-\psi _1\left(-\tfrac{1}{2} \right)  +i  \left(\mathcal{D}_{1}\psi _2(0)+\psi _1'\left(-\tfrac{1}{2} \right)\right)=0,\nonumber \\
	& \psi _3(0)-\psi _2(0) +i  \left(-\psi _3'(0)+\mathcal{D}_{1}\psi _2(0)\right)=0,\nonumber \\
		& \psi _4(0)-\psi _3(0)  +i   \left(-\mathcal{D}_{1}\psi _4(0)-\psi _3'(0)\right)=0,\nonumber \\
	& \psi _1\left(-\tfrac{1}{2} \right)-\psi _4(0) +i   \left(\psi _1'\left(-\tfrac{1}{2} \right)-\mathcal{D}_{1}\psi _4(0)\right)=0,\\[7pt]
& \varphi _2(0)-\varphi _1\left(\tfrac{1}{2} \right)  +i  \left(-\mathcal{D}_{2}\varphi _2(0)-\varphi _1'\left(\tfrac{1}{2} \right)\right)=0,\nonumber \\
& \varphi _3(0)-\varphi _2(0) +i  \left(\varphi _3'(0)-\mathcal{D}_{2}\varphi _2(0)\right)=0,\nonumber \\
& \varphi _4(0)-\varphi _3(0)  +i   \left(\mathcal{D}_{2}\varphi _4(0)+\varphi _3'(0)\right)=0,\nonumber \\
& \varphi _1\left(\tfrac{1}{2} \right)-\varphi _4(0) +i   \left(-\varphi _1'\left(\tfrac{1}{2} \right)+\mathcal{D}_{2}\varphi _4(0)\right)=0.\nonumber
\end{align}
Substituting from \eqref{ansatzB} into \eqref{FloqB} makes it possible to express the coefficients $b_3^{\pm}, a_2^{\pm}, b_4^{\pm}$ and $a_1^{\pm}$ in terms of $a_3^{\pm}, a_4^{\pm}, b_2^{\pm}$ and $b_1^{\pm}$; substituting then from \eqref{ansatzB} into \eqref{conditionB}, we get a system of eight linear equations which is solvable provided its determinant vanishes. After simple manipulations, and neglecting the inessential multiplicative factor $-2048\, k^6\, e^{i (2 \theta_2 +\theta_1 )}$, we arrive at the spectral condition
\begin{equation}\label{SCq2pos}
4 k^2-\left(k^2-1\right)^2 \cos 2 k+\left(k^2+1\right)^2 \cos 4 k+\left(k^2-1\right)^2 \Theta_2 \;\sin ^2 k=0,
\end{equation}
where the quasimomentum-dependent quantity $ \Theta_2:=\cos 2 \theta_2 +\cos \theta_1 $ ranges through the interval $[-2,2]$. Let us discuss the positive and negative part of the spectrum separately.

\subsubsection{Positive spectrum}
	\label{sect:Positq2}

\begin{figure}[h]
	\centering
	\includegraphics[scale=1.3]{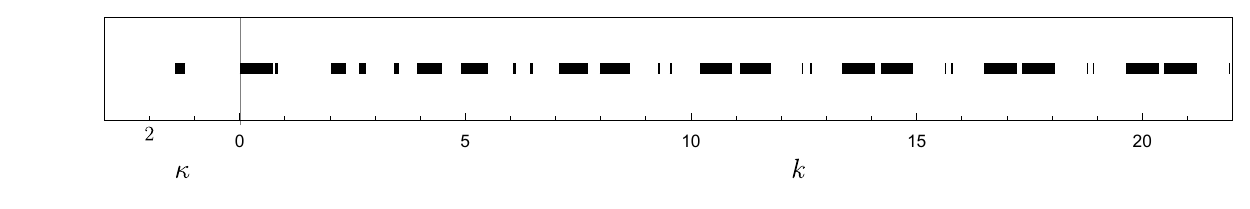}
	\caption{Spectral bands of the square lattice with the flux value of $\Phi=\pi$ per plaquette. Here and in the subsequent figures \ref{picq3} and \ref{pictotalq12}, the positive and negative spectra are simply given by $k^2$ and $-\kappa^2$, respectively.}
	\label{picq2}
\end{figure}

As in other cases where periodic quantum graphs are investigated \cite{ET18, BET22, BE22, BET20, EL19, EM17}, let us first ask whether the system can exhibit flat bands or not. This happens if the spectral condition \eqref{SCq2pos} has a solution independent of the quasimomentum components $\theta_1$ and $\theta_2\,$, or equivalently, of the quantity $\Theta_2\,$. One easily checks, however, that for $k=1$ and $k=n\pi$, $n\in\mathbb{N}$, the left-hand side of the equation \eqref{SCq2pos} reduces to $ 8 \cos ^2 2\neq 0 $ and $8 \pi ^2 n^2\neq 0$, respectively. Accordingly, there are no flat bands, and the spectrum is absolutely continuous having a band-gap structure; from condition \eqref{SCq2pos}, taking into account that $\Theta_2 \in[-2,2]$, we find that the number $k^2$ belongs to the spectral bands if and only if it satisfies the condition
\begin{equation}\label{BCq2pos}
	-2\leq \frac{-4 k^2+\left(k^2-1\right)^2 \cos 2 k-\left(k^2+1\right)^2 \cos 4 k}{\left(k^2-1\right)^2 \,\sin ^2 k}\leq 2.
\end{equation}
The band-gap pattern, containing also the negative spectrum as well, which will be discussed below, is illustrated in Fig.~\ref{picq2}.
\label{sect:Highq2}
Before turning to the negative part, let us look into the asymptotic behavior of the spectral bands in the high energy regime, $k\to\infty$. To this aim, we rewrite the spectral condition \eqref{SCq2pos} in the form
 \begin{equation}\label{q2Asy}
 	\alpha_1(k)+\frac{\alpha_2(k)}{k^2}=\mathcal{O}(k^{-4}),\end{equation}
 where
 \begin{align}
 	&  \alpha_1(k)= (-4 \cos 2 k -2+\Theta_2) \;\sin ^2 k  \;   , \nonumber\\
 	&  \alpha_2(k)= 2\left( \cos 2 k+\cos 4 k+2-\Theta_2\;\sin ^2 k \right)       .\nonumber \end{align}
Hence for large values of $k$, the solutions are close to the points where the leading term, $\alpha_1(k)$, vanishes. This gives rise to two types of spectral bands:
\begin{itemize}
\item Pairs of narrow bands in the vicinity of the roots of $\sin ^2 k$ which together with the gap between them have asymptotically constant width at the energy scale. To see that, we set $k=n\pi+\delta$ with $n\in \mathbb{N}$ and consider the limit $n\rightarrow \infty$.  Substituting it into \eqref{q2Asy} and expanding then the expression in the spectral condition at $\delta= 0$, we get in the leading order at a quadratic equation in $\delta$ which yields
\[\delta_n =\pm\frac{2 \sqrt{2}}{n\pi }\sqrt{\frac{1}{6-\Theta_2     }}+\mathcal{O}(n^{-3}) .\]
Since the band edges correspond to $\Theta_2=-2$ and $2$, the width of the bands and the gap between them are (at the energy scale) respectively determined as $\triangle E_{n,b}=2 (\sqrt{2}-1)+\mathcal{O}(n^{-2})$ and $\triangle E_{n,g}=4+\mathcal{O}(n^{-2}) $ where the subscripts $b$ and $g$ refer to the band and gap, respectively.
\item Pairs of wide bands determined by the condition $-1\leq \cos 2 k\leq 0$ corresponding to the vanishing of the bracket expression in $\alpha_1(k)$.
\end{itemize}

\smallskip

\noindent The different character of those bands is illustrated in Fig.~\ref{picq2}. It is also seen in the probability that an energy value belongs to the spectrum for a randomly chosen value of the momentum $k$,
\begin{equation}\label{probsigma}
	P_{\sigma}(H):=\lim_{K\to\infty} \frac{1}{K}\left|\sigma(H)\cap[0,K]\right|.
\end{equation}
This quantity was introduced by Band and Berkolaiko \cite{BB13} who demonstrated its universality -- meaning its independence on graph edge lengths as long as they are incommensurate -- in periodic graphs with Kirchhoff coupling; the claim was recently extended to graphs with the coupling considered here \cite{BE22,BEL22}. For equilateral graphs which we deal with here the universality makes naturally no sense but the probability \eqref{probsigma} can be useful to compare our results to those concerning discrete magnetic Laplacians of \cite{Sh94}. The width of the narrow bands at the momentum scale $\triangle k_{n,b}=\frac{ \sqrt{2}-1}{n\pi}+\mathcal{O}(n^{-3})$, and as a consequence, their contribution  to the probability \eqref{probsigma} is zero. For the wide bands, on the other hand, it obviously equals $\tfrac12$.

\subsubsection{Negative spectrum}
	\label{sect:Negatq2}

As we have mentioned, to find the negative spectrum, it is only needed to replace $k$ by $i\kappa$ in \eqref{SCq2pos} and \eqref{BCq2pos} which, respectively, leads to the following spectral and band conditions
	\begin{equation}\label{SCq2neg}
		 -4 \kappa ^2- (\kappa ^2+1 )^2 \cosh 2 \kappa + (\kappa ^2-1 )^2 \cosh 4 \kappa - (\kappa ^2+1 )^2 \;\Theta_2\;\sinh ^2 \kappa=0,
	\end{equation}
	\begin{equation}\label{BCq2neg}
		-1\leq    \frac{ \kappa ^4-6 \kappa ^2+2 (\kappa ^2-1)^2 \cosh 2 \kappa -4 \kappa ^2\, \text{csch}^2 \kappa +1}{ (\kappa ^2+1 )^2}    \leq 1.
	\end{equation}
More precisely, a negative eigenvalue $-\kappa^2$ belongs to a spectral band if the positive number $\kappa$ satisfies the band condition \eqref{BCq2neg}. As Fig.~\ref{picq2} illustrates, there is no flat band which is obvious from the spectral condition \eqref{SCq2neg} in which the coefficient of $\Theta_2$ in the last, the only quasimomentum dependent term is nonzero. Concerning the number of negative bands, let us first recall that the Hamiltonian of a star graph with $N\ge 3$ semi-infinite edges and the coupling given by the matrix $U_v$ (see \eqref{eq:uv}) at the central vertex has a nonempty discrete spectrum in the negative part in which the eigenvalues are by \cite{ET18} equal to
\begin{equation}\label{Neg,Eig,StarG}
	E=-\tan^{2}\frac{m\pi}{N},
\end{equation}
with $m$ running through $ 1,\cdots,[\tfrac{N}{2}] $ for odd $N$ and $1,\cdots,[\tfrac{N-1}{2}]$ for even $N$; their number coincides with the number of eigenvalues of the matrix $U_v$ with positive imaginary part. Since the unit-flux cell contains for $\Phi=\pi$ two vertices of degree four, the corresponding matrix $U_v$ in each of them has only one negative eigenvalue in the upper complex halfplane, and consequently, in accordance with Theorem 2.6 of \cite{BET22} the negative spectrum cannot have more than two bands.

However, as can be seen in Fig.~\ref{picq2}, in reality there is only one negative band which can be checked by inspecting the band condition \eqref{BCq2neg}. Denoting the function inside the inequality by $f(\kappa)$, we find that each condition $f(\kappa)=\pm1$ can have only one solution for $\kappa>0$. Indeed, consider first the condition $f(\kappa)=-1$ which, after simple manipulations, can be rewritten in the factorized form
\begin{equation}\label{proofneg1}
	(\kappa -\coth \kappa ) (\kappa -\tanh \kappa ) (\sinh \kappa +\kappa  \cosh \kappa ) (\kappa  \sinh \kappa +\cosh \kappa )\;\text{csch}^2\kappa=0\,,
\end{equation}	
where we have divided $f(\kappa)+1$ by $4\,\sinh \kappa\,\cosh \kappa > 0 $ and multiplied by $(\kappa^2 + 1)^2$. It is easy to check that only the expression in the first bracket in \eqref{proofneg1} can be zero since it is monotonically increasing (with the first derivative, $\coth ^2\kappa$, positive) on the interval $\kappa\in(0,\infty)$ ranging from $-\infty$ to $+\infty$, which implies that it has only one root on the domain. The second term, $\kappa -\tanh \kappa$, cannot be zero since it is also monotonically increasing on the domain (with the first derivative $\tanh ^2\kappa >0$) but ranging from $0$ to $+\infty$; the expressions in the last two brackets are obviously nonzero for $\kappa>0$, needless to say that the last term, $\text{csch}^2\kappa$, cannot give a solution since the left hand side of \eqref{proofneg1} tends to infinity as $\kappa\to\infty$. Now, let us pass to the condition $f(\kappa)=1$ which, after manipulations, can be rewritten as
$$ (\coth ^2 \kappa +1)\lambda(\kappa ) =0; \qquad  \lambda(\kappa ):=  (\kappa ^2-1 )^2 \cosh 2\kappa - (\kappa ^2+1 )^2, $$
in which $\coth ^2 \kappa +1$ is obviously nonzero; moreover, we see that $\lambda(\kappa )$ is negative for $\kappa\in(0,1]$ in view of that $0\leq(\kappa ^2-1 )^2 \cosh 2\kappa<1$ while $(\kappa ^2+1 )^2>1$, note that $\cosh 2\kappa>1\,$ for $\kappa>0$. Hence, it suffices to inspect the interval $\kappa\in(1,\infty)$; to this aim, we rewrite the condition $\lambda(\kappa)=0$ in the new form $\xi(\kappa):=\cosh 2 \kappa-(\tfrac{\kappa ^2+1}{\kappa ^2-1})^2=0$ from which we have $\xi'(\kappa)=2 \sinh 2\kappa+\frac{8  (\kappa ^3+\kappa  )}{ (\kappa ^2-1 )^3} >0\,$ for $\kappa>1$, implying that $\xi(\kappa)$ is monotonically increasing on the domain. On the other hand, we have $\lim_{\kappa \to 1}\xi(\kappa )=-\infty$ and  $\lim_{\kappa \to \infty} \xi(\kappa )=+\infty\,$; this, together with the monotonicity of $\xi(\kappa)$, confirms that it can have only one root on the mentioned domain which concludes the claim.

\subsection{The case $\Phi=2\pi\tfrac p3$}
	\label{sect:fluxq3}

Let us consider next the case when the flux value per plaquette is $\Phi= 2\pi\tfrac{p}{3}$, where $p=\{1,2\}$; the elementary cell now contains three vertices of degree four. The spectral condition can be derived in a similar way as in Sec.~\ref{sect:fluxpi} by employing the appropriate Ans\"atze and matching them at each vertex as we did when deriving \eqref{SCq2pos}; note that in this case one has to also consider the quasi-derivative $\mathcal{D}_3:= \frac{d}{dy}- 3i B $ referring  the vertical edge passing through the third vertex. Seeking non-trivial solutions of the corresponding system of twelve linear equations, we arrive at the spectral condition
\begin{equation}\label{SCq3pos}
	g(k)+ (k^2-1 )^3 \;\Theta_3\;\sin ^3 k=0,
\end{equation}
where $ \Theta_3:=\cos 3 \theta_2 +\cos \theta_1 \in[-2,2]$ and
\begin{align*}\label{gkappa}
   	g(k):=&\;\; 6  (k^2-1 )^2  (k^2+1 ) \sin k\; \cos ^3 k- (k^2+1 )^3 \sin 6 k  \\
	&   -8 k \Big( (k^4+6 k^2+1 ) \cos 2 k- (k^2-1)^2 \Big) \sin ^3 \frac{\pi  p}{3} \, \cos  \frac{\pi  p}{3}     \\
	&    -3  (k^2+1 )  \Big( (k^2-1 )^2 \cos 2 k- (k^2+1 )^2\Big) \sin 2 k\;\cos  \frac{2 \pi  p}{3}.
\end{align*}
	
	\subsubsection{Positive spectrum}
	\label{sect:Positq3}
	
\begin{figure}[h]
		\centering
		\includegraphics[scale=1.3]{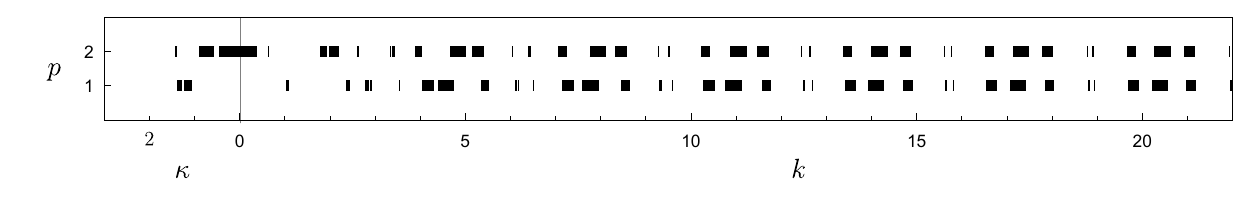}
		\caption{Spectral bands of the square lattice with the flux per plaquette $\Phi=2\pi\frac p3\,$, $p=\{1,2\}$.}
		\label{picq3}
	\end{figure}

As in Sec.~\ref{sect:Positq2}, we ask first whether the spectral condition \eqref{SCq3pos} can give rise to flat bands or not; the values to explore are again $k=1$ and $k=n\pi$ with $n\in\mathbb{N}$, for which the quasimomentum-dependent part of the spectral condition vanishes. Evaluating the function $g(k)$ at the corresponding energies, we arrive at the expressions $ -8 \sin 6-64 \cos 2 \sin ^3 \frac{\pi  p}{3}  \cos  \frac{\pi  p}{3} +24 \sin 2 \cos  \frac{2 \pi  p}{3}$ and $-64 \pi ^3 n^3 \sin ^3 \frac{\pi  p}{3}  \cos  \frac{\pi  p}{3}$, respectively; one easily checks that both the expressions are nonzero for $p=\{1,2\}$, so that there is no flat band. The spectrum is thus absolutely continuous having a band-gap structure, as illustrated in Fig.~\ref{picq3}. More explicitly, using \eqref{SCq3pos} and taking into account the range of the quasimomentum-dependent quantity $\Theta_3$, we find that an energy $k^2$ belongs to a spectral band if and only if it satisfies the condition
\begin{equation}\label{BCq3pos}
	-2\leq  \frac{g(k)}{- (k^2-1)^3 \;\sin ^3 k}  \leq 2.
\end{equation}
As before, we are interested in the asymptotic behavior of the bands in the high energy regime, $k\to\infty$. To this aim, we rewrite the spectral condition \eqref{SCq3pos} in the form
	\begin{equation}\label{q3Asy}
		\beta_1(k)+\frac{\beta_2(k)}{k}+\frac{\beta_3(k)}{k^2}+\frac{\beta_4(k)}{k^3}=\mathcal{O}(k^{-4}),\end{equation}
	where
	\begin{align}
		&  \beta_1(k)=\Big(\Theta_3+6 \cos  \big(k+\frac{2 \pi  p}{3} \big)+6 \cos  \big(k-\frac{2 \pi  p}{3} \big)+18 \cos k+8 \cos 3k \Big) \sin ^3 k   \;   , \nonumber\\
		&  \beta_2(k)= 8 \sin ^2 \frac{\pi  p}{3}\;  \sin \frac{2 \pi  p}{3} \;   \sin ^2 k    \;   , \nonumber\\
		&  \beta_3(k)=\frac{3}{2} (6 \sin 2 k+\sin 4 k) \cos \frac{2 \pi  p}{3}-3 \Theta_3 \sin ^3 k-3 \sin 6 k-6 \sin k \,\cos ^3 k \;   , \nonumber\\
		&  \beta_4(k)=-16 (3 \cos 2 k+1) \sin ^3 \frac{\pi  p}{3} \, \cos  \frac{\pi  p}{3} . \end{align}
Seeking the solution in the vicinity of the points where the leading term, $\beta_1(k)$, vanishes, we find again two types of spectral bands:

\begin{itemize}
	\item Series of three narrow bands in the vicinity of the roots of $\sin ^3 k\,$ which, again, have an asymptotically constant width on the energy scale as $k\rightarrow \infty$. To estimate the width of these bands, as in the previous case, we set $k=n\pi+\delta$ with $n\in \mathbb{N}$ and consider the limit $n\rightarrow \infty$; substituting it into \eqref{q3Asy} and expanding then the resulting equation at $\delta= 0$, we get in the leading order at a cubic equation in $\delta$ from which we obtain three solutions for $\delta$ in the form of Cardano's formula that are asymptotically of the form
	\[\delta_{n,j}= \frac{ \mathcal{G}(p, \Theta_3)}{ n  }+\mathcal{O}(n^{-3}),\qquad j=1,2,3 .\]
Then, taking into account that the band edges correspond to $\Theta_3=-2$ and $2$, the width of these three bands, $\triangle E_{n,j}(p)$, $j=1,2,3$, at the energy scale is obtained as
\begin{align*}
	&   \triangle E_{n,1}(1)=\triangle E_{n,3}(2)= \frac{1}{\sqrt{3}}  +\mathcal{O}(n^{-2})  \;   , \\
	&  \triangle E_{n,2}(1)=  \triangle E_{n,2}(2)= \frac{1}{22} \left(3 \sqrt{3}-11 \sqrt{19}+48\right)  +\mathcal{O}(n^{-2})  \;   , \\
	&  \triangle E_{n,3}(1)=\triangle E_{n,1}(2)= \frac{1}{22} \left(3 \sqrt{3}+11 \sqrt{19}-48\right) +\mathcal{O}(n^{-2}) . \end{align*}
The results, as expected, coincide with the narrow bands pattern in Fig.~\ref{picq3} confirming that these bands are asymmetric with respect to a swap of $p$ and $q-p$; for more details, see the fourth bullet point in Sec.~\ref{sect:fluxpq12}.
	\item Series of three wide bands determined by the condition
\begin{equation}\label{highBandConq3}
-1\leq \;-3 \cos \big(k+\frac{2 \pi  p}{3}\big)-3 \cos \big(k-\frac{2 \pi  p}{3}\big)-9 \cos k-4 \cos3k\;\leq 1,
\end{equation}
referring to the vanishing of the `large' bracket in $\beta_1(k)$, taking into account that $\Theta_3\in[-2,2]$; note that for $q=3$ one easily checks that the function in the inequality is the same for $p=1,2$ and condition \eqref{highBandConq3} simplifies to
\begin{equation}\label{highBandConq3new}
	-1\leq \;  -6 \cos k-4 \cos 3k\;\leq 1.
\end{equation}
\end{itemize}

\smallskip

\noindent The narrow bands again do not contribute to \eqref{probsigma}. As for the wide ones, it is easy to compute on a single period of the function inside the inequality \eqref{highBandConq3new} the ratio of the sum of intervals of $k$ satisfying the condition to the period; this shows that the probability \eqref{probsigma} of belonging to the spectrum for a randomly chosen value of $k$ equals $P_{\sigma}(H)=-\tfrac13+\tfrac{4}{\pi}\, \arctan \sqrt{6-\sqrt{33}} \approx 0.262498$. The result is independent of $p$; in the following we will see that this is no longer true for higher values of $q\,$.

	\subsubsection{Negative spectrum}
	\label{sect:Negatq3}

The corresponding spectral condition is again obtained by replacing the momentum variable $k$ in \eqref{SCq3pos} and \eqref{BCq3pos} by $i\kappa$. The spectrum is absolutely continuous; there is no flat band as one can check in a way analogous to that of Sec.~\ref{sect:Negatq2}. The unit-flux cell now contains three vertices so in accordance with Theorem~2.6 of \cite{BET22} the negative spectrum cannot have more than three spectral bands; as we see in Fig.~\ref{picq3}, this happens for $p=2$ while for $p=1$ we have only two negative bands.

	\subsection{The case $\Phi=2\pi\frac pq$ with $q=\{2,3,...,12\}$}
	\label{sect:fluxpq12}

After dealing with the two simplest cases, we pass to the situation with the flux values $\Phi=2\pi\frac pq$ for all the coprime ratios $\frac pq$ with $q\in\{2,...,12\}$ and $p=1,...,q-1\,$. As we have already mentioned, the higher the $q$ becomes, the more complicated the spectral condition is; we proceeded here to the limit of what was computationally manageable, reaching it at systems of 48 linear equations. The scheme remains the same; the spectral condition takes generally the form
\begin{equation}\label{GenSCq}
h(k;p,q)+ (k^2-1 )^q  \;\Theta_q\;\sin ^q k=0,
\end{equation}
where the quasimomentum-dependent quantity $\Theta_q:=\cos q\theta_2 +\cos \theta_1 $ ranges again through the interval $[-2,2]$. While the latter is simple and depends as in Secs.~\ref{sect:fluxq3}	and ~\ref{sect:fluxpi} on $q$ only, we have not been able to find a general expression of the function $h(k;p,q)$ for any $p$ and $q$.

\begin{figure}[htb!]
		\centering
		\includegraphics[scale=0.6]{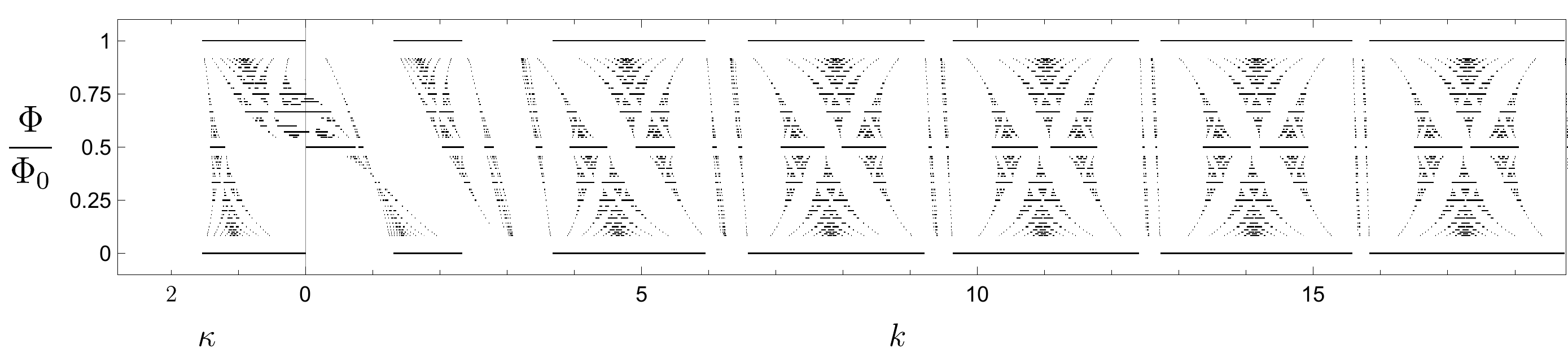}
		\caption{Spectrum of the square lattice of unit edge length for the flux ratio per plaquette $\frac{\Phi}{\Phi_0}=\frac pq$ with $q\in\{2,...,12\}$ and $p=1,...,q-1$. At the top and bottom the spectral bands of $\Phi \in 2\pi\mathbb{Z}$ corresponding to the non-magnetic case \cite{ET18} are shown.}
	\label{pictotalq12}
	\end{figure}

\begin{figure}[htb!]
	\centering
	\includegraphics[scale=1.2]{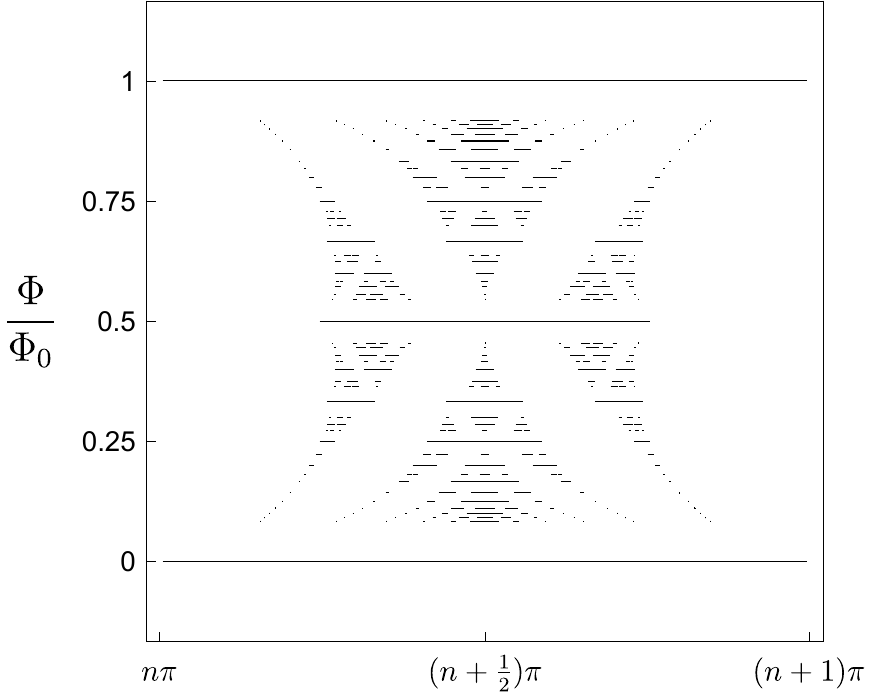}
	\caption{The asymptotic shape of the butterfly part of the spectrum. At the top and bottom the spectral bands of the non-magnetic case are again shown.}
	\label{pichigh}
\end{figure}

\begin{figure}[htb!]
	\centering
	\includegraphics[scale=0.60]{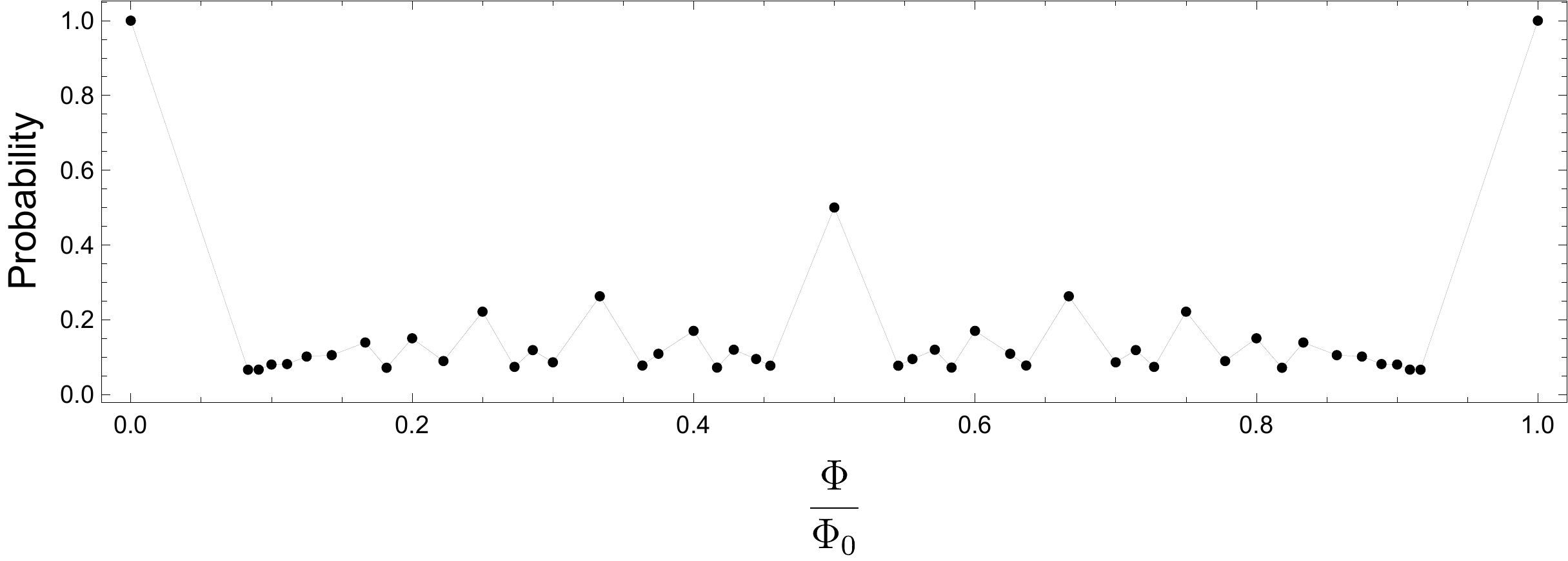}
	\caption{ The probability \eqref{probsigma} of belonging to the spectrum for a randomly chosen value of $k$. To make the pattern more visible, we join the points referring to the adjacent values of $\frac{\Phi}{\Phi_0}$.}
	\label{picprob}
\end{figure}

\begin{figure}[htb!]
	\centering
	\hspace{2em}\includegraphics[scale=0.8]{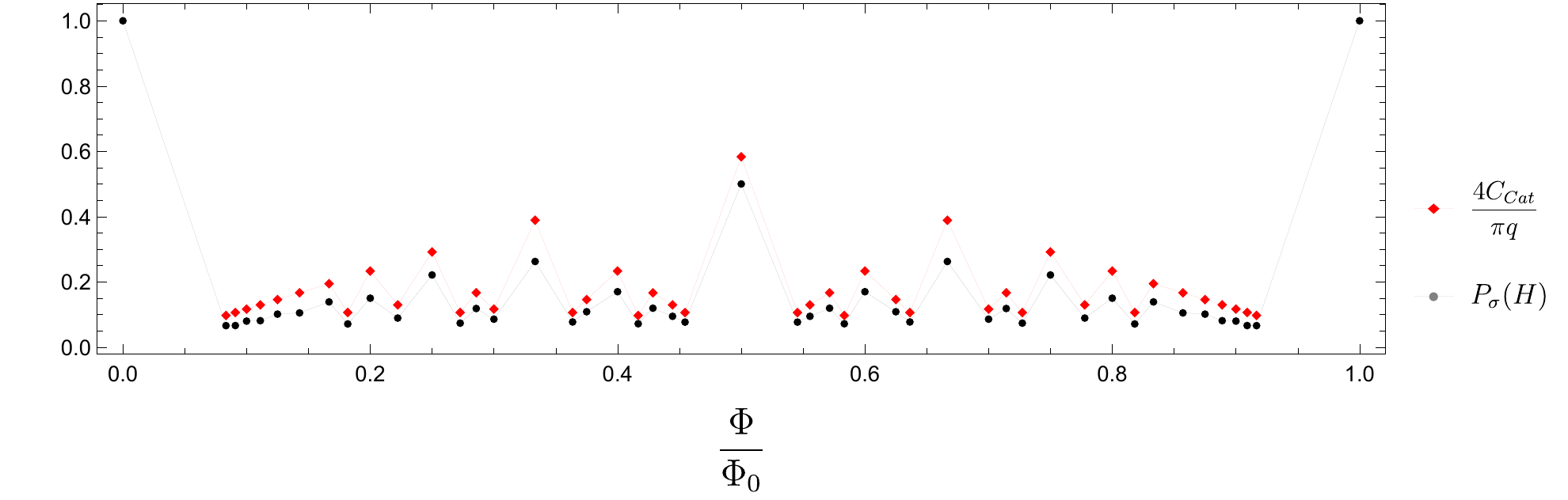}
	\caption{Comparison of \eqref{probsigma} to the Thouless conjecture values indicated by the red diamonds.}
	\label{measure-comp}
\end{figure}

One can check that $h(k;p,q)$ does not vanish at $k=1$ and $k=n\pi$ so that the spectrum is absolutely continuous having a band-and-gap character. The results of the computation for the considered flux values given in Fig.~\ref{pictotalq12} show an intricate spectral pattern:
\begin{itemize}	
 \item Given the fact that $ \Theta_q \in[-2,2]$, away from $k=1$ and $k=n\pi$ a positive number $k^2$ belongs to the spectrum if and only if
 \begin{equation}\label{BCqpos}
 	-2\leq  \frac{h(k;p,q)}{(k^2-1 )^q \;\sin ^q k}  \leq 2\,;
 \end{equation}
 given the increasingly complicated form of the function $h(\cdot;p,q)$ the number of bands becomes larger with increasing $q$ which motivates one to conjecture that the spectrum could be \emph{fractal}, in fact a \emph{Cantor set}, for $\Phi\not\in2\pi\mathbb{Q}$. As the momentum $k$ grows, the dominating part of the spectrum -- for the sake of brevity we label it for obvious reasons as \emph{butterfly} -- can be found within the spectral bands of the non-magnetic lattice; recall that in that case the bands dominate the spectrum \cite{ET18}.
 \item There are also `narrow' bands in the gaps of the non-magnetic spectrum. In the same way as for $q=2,3$ one can check that they cover intervals of order $\mathcal{O}(n^{-1})$ at the momentum scale, hence the weight of this part of the spectrum -- let us call it \emph{non-butterfly}-- diminishes as $k$ increases. Its components remain nevertheless visible being of asymptotical constant width at the energy scale, separated by linearly blowing up butterfly patterns. The numerical results also show that the number of bands in these parts increases with growing $q$, and one can conjecture that this part of the spectrum would again have a Cantor character for $\Phi\not\in2\pi\mathbb{Q}$.
 \item The negative spectrum also has a band-gap structure coming from condition \eqref{BCqpos} in which $k$ is replaced by $i\kappa$ with $\kappa>0$; in accordance with Theorem 2.6 of \cite{BET22} it cannot have more than $q$ spectral bands for the flux value $\Phi=2\pi\frac pq$ per plaquette.
 \item Looking at Fig.~\ref{pictotalq12}, we see that the vertex coupling influences the spectral pattern strongly in the lowest parts of spectrum, including the negative one; at higher energies the effect of the magnetic field prevails. The vertex effect never disappears fully, though. In particular, the non-butterfly parts of the spectrum, small as it may be, come in pairs of band clusters, which -- in contrast to the rest of the spectrum -- are, even at high energies, visibly asymmetric with respect to a swap of $p$ and $q-p$, which is equivalent to flipping the field direction.
 \item In the high energy regime, $k\to\infty$, one can rewrite the spectral condition \eqref{GenSCq} in the polynomial form
 \begin{equation}\label{GenSCqhigh}
 	k^{2q}\,\Big(\Theta_q + w(k; p,q) \Big) \sin ^q k  +\mathcal{O}(k^{2q-1})=0,
 \end{equation}
where $w(\cdot; p,q)$ is a periodic function, generalizing the functions $\frac{\alpha_1(k)}{\sin ^2 k } - \Theta_2,\,\frac{\beta_1(k)}{\sin ^3 k }  - \Theta_3$ above, which means that the butterfly part is asymptotically $\pi$-periodic in the momentum variable. We plot the pattern obtained from the requirement of the large bracket vanishing in \eqref{GenSCqhigh} in Fig.~\ref{pichigh} for one interval, $n\pi\leq k \leq (n+1)\pi$. Despite the computational restriction on the value of $q$ we see Hofstadter's pattern clearly emerging.
 \item In view of the asymptotic periodicity and the fact that non-butterfly part becomes negligible as $n\to\infty$, the probability $P_{\sigma}(H)$ given by \eqref{probsigma} coincides with the Lebesgue measure of the band spectrum normalized to one on the interval $(n\pi,(n+1)\pi)$. We plot this quantity in dependence of the considered flux values in Fig.~\ref{picprob}; it decreases with the growing $q$ in accordance with the expectation that for $\Phi\not\in2\pi\mathbb{Q}$ the spectrum has measure zero. Recall further that while the `total bandwidth' of the rational Harper operator is in general not known, for increasingly complicated coprime ratios we have the Thouless conjecture \cite{Th83, HK95} which states that
 \begin{equation}\label{Thouless}
 	\lim_{\scriptsize \begin{array}{c} q\to\infty \\ p\wedge q=1 \end{array}} q\big|\sigma\big(\Phi=2\pi\tfrac pq\big)\big| = \frac{16\, C_\mathrm{Cat}}{\pi}
 \end{equation}
with the Catalan constant $C_\mathrm{Cat} = \sum_{n\in\mathbb{N}} (-1)^n (2n+1)^{-2} \approx 0.9159...$. We can compare this claim with \eqref{probsigma}. Having in mind that the standard interval on which the Hofstadter's butterfly is plotted is $[-2,2]$, the normalized measure values of $\frac{4C_\mathrm{Cat}}{\pi q}$ are according to Fig.~\ref{measure-comp} not far from $P_\sigma(H)$ even for the relatively small values of $q$ we consider here.
\end{itemize}


	\clearpage
\subsection*{Data availability statement}

Data are available in the article.

\subsection*{Conflict of interest}

The authors have no conflict of interest.


\subsection*{Acknowledgments}
M.B. and J.L. were supported by the Czech Science Foundation within the project 22-18739S. The work of P.E. was supported by the Czech Science Foundation within the project 21-07129S and by the EU project CZ.02.1.01/0.0/0.0/16\textunderscore 019/0000778.


\end{document}